# Compact electron acceleration and bunch compression in THz waveguides

Liang Jie Wong,[1,*] Arya Fallahi,[2] and Franz X. Kärtner[1,2,3]

[1] *Department of Electrical Engineering and Computer Science and Research Laboratory of Electronics, Massachusetts Institute of Technology, 77 Massachusetts Avenue, Cambridge, MA, 02139, USA*
[2] *Center for Free-Electron Laser Science, DESY, Notkestraße 85, D-22607 Hamburg, Germany*
[3] *Department of Physics, University of Hamburg, Notkestraße 85, D-22607 Hamburg, Germany*
[*]*ljwong@mit.edu*

**Abstract:** We numerically investigate the acceleration and bunch compression capabilities of 20 mJ, 0.6 THz-centered coherent terahertz pulses in optimized metallic dielectric-loaded cylindrical waveguides. In particular, we theoretically demonstrate the acceleration of 1.6 pC and 16 pC electron bunches from 1 MeV to 10 MeV over an interaction distance of 20mm, the compression of a 1.6 pC 1 MeV bunch from 100 fs to 2 fs (50 times compression) over an interaction distance of about 18mm, and the compression of a 1.6 pC 10 MeV bunch from 100 fs to 1.61 fs (62 times) over an interaction distance of 42 cm. The obtained results show the promise of coherent THz pulses in realizing compact electron acceleration and bunch compression schemes.

**References and links**

1. E. Esarey, C. B. Schroeder, and W. P. Leemans, "Physics of laser-driven plasma-based electron accelerators," Rev. Mod. Phys. **81,** 1229–1285 (2009).
2. T. Plettner, R. L. Byer, E. Colby, B. Cowan, C. M. S. Sears, J. E. Spencer, and R. H. Siemann, "Visible-laser acceleration of relativistic electrons in a semi-Infinite vacuum," Phys. Rev. Lett. **95,** 134801 (2005).
3. G. Malka, E. Lefebvre, and J. L. Miquel, "Experimental observation of electrons accelerated in vacuum to relativistic energies by a high-intensity laser," Phys. Rev. Lett. **78,** 3314–3317 (1997).
4. A. Karmakar and A. Pukhov, "Collimated attosecond GeV electron bunches from ionization of high-Z material by radially polarized ultra-relativistic laser pulses," Laser Part. Beams **25,** 371–377 (2007).
5. S. Kawata, Q. Kong, S. Miyazaki, K. Miyauchi, R. Sonobe, K. Sakai, K. Nakajima, S. Masuda, Y. K. Ho, N. Miyanaga, J. Limpouch, and A. A. Andreev, "Electron bunch acceleration and trapping by ponderomotive force of an intense short-pulse laser," Laser Part. Beams **23,** 61–67 (2005).
6. A. Mizrahi and L. Schächter, "Optical bragg accelerators," Phys. Rev. E **70,** 016505 (2004).
7. Y. C. Huang, D. Zheng, W. M. Tulloch, and R. L. Byer, "Proposed structure for a crossed-laser beam, GeV per meter gradient, vacuum electron linear accelerator," Appl. Phys. Lett. **68,** 753–755 (1996)
8. T. Plettner, P. P. Lu, and R. L. Byer, "Proposed few-optical cycle laser-driven particle accelerator structure," Phys. Rev. ST Accel. Beams **9,** 111301 (2006).
9. W. Gai, P. Schoessow, B. Cole, R. Konecny, J. Norem, J. Rosenzweig, and J. Simpson, "Experimental demonstration of wake-field effects in dielectric structures," Phys. Rev. Lett. **61,** 2756–2758 (1988).
10. G. Andonian, D. Stratakis, M. Babzien, S. Barber, M. Fedurin, E. Hemsing, K. Kusche, P. Muggli, B. O'Shea, X. Wei, O. Williams, V. Yakimenko, and J. B. Rosenzweig, "Dielectric wakefield acceleration of a relativistic electron beam in a slab-symmetric dielectric lined waveguide," Phys. Rev. Lett. **108**, 244801 (2012).
11. S. Antipov, C. Jing, A. Kanareykin, J. E. Butler, V. Yakimenko, M. Fedurin, K. Kusche, and W. Gai, "Experimental demonstration of wakefield effects in a THz planar diamond accelerating structure," Appl. Phys. Lett. **100** 132910 (2012).
12. M. C. Hoffmann and J. A. Fülöp, "Intense ultrashort terahertz pulses: generation and applications," J. Phys. D **44,** 083001 (2011).
13. J. A. Fülöp, L. Pálfalvi, G. Almási, and J. Hebling, "Design of high-energy terahertz sources based on optical rectification," Opt. Express **18,** 12311–12327 (2010).

14. J. A. Fülöp, L. Pálfalvi, M. C. Hoffmann, and J. Hebling. "Towards generation of mJ-level ultrashort THz pulses by optical rectification." Opt. Express **19**, 15090–15097 (2011).
15. S.-W. Huang, E. Granados, W. R. Huang, K.-H. Hong, L. E. Zapata and F. X. Kärtner, "High conversion efficiency, high energy terahertz pulses by optical rectification in cryogenically cooled lithium niobate," Opt. Lett. **38**, 796-798 (2013)
16. J. Hebling, J. A. Fülöp, M. I. Mechler, L. Pálfalvi, C. Tőke, and G. Almási, "Optical manipulation of relativistic electron beams using THz pulses," arXiv:1109.6852.
17. R. B. Yoder and J. B. Rosenzweig, "Side-coupled slab-symmetric structure for high-gradient acceleration using terahertz power," Phys. Rev. ST Accel. Beams **8,** 111301 (2005).
18. G. Sciaini and R. J. D. Miller. "Femtosecond electron diffraction: heralding the era of atomically resolved dynamics," Rep. Prog. Phys. **74** 096101 (2011).
19. H. Yoneda, K. Tokuyama, K. Ueda, H. Yamamoto, and K. Baba, "High-power terahertz radiation emitter with a diamond photoconductive switch array," Appl. Optics **40**, 6733–6736 (2001).
20. V. V. Kubarev, "Optical properties of CVD-diamond in terahertz and infrared ranges," Nucl. Instr. and Meth. A **603,** 22–24 (2001).
21. J. H. Kim, J. Han, M. Yoon, and S. Y. Park, "Theory of wakefields in a dielectric-filled cavity," Phys. Rev. ST Accel. Beams **13**, 071302 (2010).
22. L. D. Landau and E. M. Lifshitz, *The Classical Theory of Fields*, 4th ed. (Oxford 1987).
23. J. D. Jackson, *Classical Electrodynamics*, 2nd ed. (Wiley, 1975).
24. *GPT User Manual*, Pulsar Physics.
25. W. H. Press, S. A. Teukolsky, W. T. Vetterling and B. P. Flannery, *Numerical Recipes in C*, 2nd ed. (Cambridge University, 1992, pp. 714-720).
26. T. F. Chan, G. H. Golub, and R. J. LeVeque, "Algorithms for computing the sample variance: Analysis and recommendations," The American Statistician **37,** 242-247 (1983)
27. P. Yeh, A. Yariv, and E. Marom, "Theory of Bragg fiber*,"* J. Opt. Soc. Am. **68,** 1196–1201 (1978).
28. G. Gallot, S. P. Jamison, R. W. McGowan, and D. Grischkowsky. "Terahertz waveguides." J. Opt. Soc. Am. B **17**, 851–863 (2000).
29. G. Chang, C. J. Divin, C.-H. Liu, S. L. Williamson, A. Galvanauskas, and T. B. Norris. "Generation of radially polarized terahertz pulses via velocity-mismatched optical rectification." Opt. Lett. **32**, 433–435 (2007).
30. S. Winnerl, B. Zimmermann, F. Peter, H. Schneider, and M. Helm. "Terahertz Bessel-Gauss beams of radial and azimuthal polarization from microstructured photoconductive antennas." Opt. Express **17**, 1571–1576 (2009).
31. T. Grosjean, F. Baida, R. Adam, J.-P. Guillet, L. Billot, P. Nouvel, J. Torres, A. Penarier, D. Charraut, and L. Chusseau. "Linear to radial polarization conversion in the THz domain using a passive system." Opt. Express **16**, 18895–18909 (2008).
32. J. Yang, N. Naruse, K. Kan, T. Kondoh, Y. Yoshida, and K. Tanimura, "Femtosecond electron guns for ultrafast electron diffraction," *International Particle Accelerator Conference* (IPAC'2012), paper FRXBB01.

## 1. Introduction

The desire to realize electron acceleration schemes that can surpass the approximately 50 MeV/m maximum acceleration gradient of conventional radio-frequency (RF) technology has spurred much research into the use of alternative regions of the electromagnetic spectrum. Methods that have been investigated include laser-induced plasma acceleration [1], vacuum acceleration [2-5] using optical pulses and dielectric-based acceleration. Dielectric-based acceleration is achieved either by an external optical laser source [6-8] or by the wakefields of another electron bunch (i.e. dielectric wakefield accelerator) [9-11]. With the advent of efficient coherent THz pulse generation techniques [12-15], forays have also been made into the acceleration of electrons in vacuum [16] and in waveguides [17] by coherent THz pulses.

This paper demonstrates the capabilities of waveguides optimized for acceleration and/or compression of relativistic electron bunches by coherent THz pulses. The relativistic few-femtosecond pico-Coulomb electron bunch achieved in the bunch compression scheme has applications in single-shot few-femtosecond electron diffraction [18]. We choose to study dielectric-loaded cylindrical metallic waveguides for their ease of manufacturing and theoretical evaluation. The THz frequency range is chosen as the operation range because it appears to strike a compromise between the large wavelength and low acceleration gradient (due to breakdown limitations) of RF radiation and the small wavelength but high acceleration gradient of optical radiation. Note that a higher acceleration gradient is more favorable for bunch compression and acceleration, but space-charge effects make it difficult to confine a bunch of substantial charge well within a half-cycle if the wavelength is too small. The absence of plasma in a vacuum-core waveguide scheme precludes problems associated with the inherent instability of laser-plasma interactions. Although using a guiding structure leads to intensity limitations, it also increases acceleration efficiency due to a smaller driving energy required and a larger interaction distance.

The high thermal conductivity and breakdown properties of chemical-vapor-deposited diamond at THz frequencies are well-recognized, and has led to its use in waveguides for wakefield acceleration [11] and other applications involving intense terahertz radiation [19]. For this reason, we use diamond for the dielectric throughout this study and assume a relative dielectric constant of $\varepsilon_r$ = 5.5 [20]. We employ the fundamental transverse-magnetic waveguide mode ($TM_{01}$ mode) because every field component in this mode vanishes on axis except for the z-directed electric field, so an electron bunch close to the axis will be accelerated mainly in the forward direction.

Figure 1 illustrates an example of concurrent compression and acceleration of an electron bunch in our scheme. We present this example before any technical discussion to give some preliminary intuition of the electrodynamics that ensues when a 1 MeV electron bunch (obtained, for instance, from an RF gun) is injected into a coherent THz pulse propagating in a dielectric-loaded cylindrical metal waveguide.

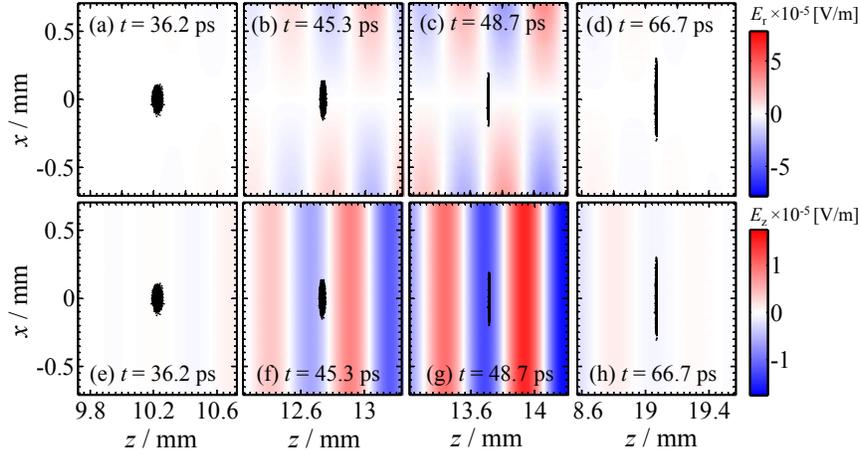

Fig. 1. Illustration of electron bunch acceleration and compression by a $TM_{01}$ coherent THz pulse in a dielectric-loaded (diamond) cylindrical metal waveguide: The 8-cycle pulse is centered at 0.6 THz, with group velocity 0.399c and phase velocity c. The 1.6 pC-bunch has an initial mean kinetic energy of 1 MeV. Steps of the bunch evolution include: (a) arriving at the rear of the pulse, (b) slipping through an accelerating and compressing quarter-cycle, (c) maximum longitudinal compression and (d) transverse and longitudinal expansion as the electron bunch emerges from the head of the pulse. Each black dot indicates a macro-particle, with 1000 macro-particles used in the simulation. The color maps in (a)-(d) show the value of $E_r$ in the y = 0 plane. (e)-(h) is identical to (a)-(d) respectively, except that the color maps show $E_z$ instead of $E_r$.

Note that the work pursued here differs from the study presented in [17], which discusses the design of a uniformly-accelerating 100 MeV/m coherent THz pulse-driven waveguide accelerator. Here, we study the acceleration as well as bunch compression capabilities of a coherent THz pulse of finite duration. Moreover, the presented simulation results for coherent THz pulse-driven acceleration and compression cannot be taken for granted or inferred by scaling the results from studies at optical or RF frequencies, because of the non-negligible impact of space-charge.

In Sec. 2 and 3, we furnish a technical discussion of the equations upon which our model rests. In Sec. 4, we demonstrate the acceleration of a 1.6 pC electron bunch from a kinetic energy of 1 MeV to about 10 MeV over an interaction distance of about 20mm, using a 20mJ pulse centered at 0.6 THz in a dielectric-loaded metallic waveguide. The implications of using an arbitrarily distant injection point, as well as the prospects of dielectric breakdown and thermal damage for our optimized design are also analyzed. In Sec. 5, we investigate the acceleration of 16 pC and 160 pC 1 MeV electron bunches. In Sec. 6, we optimize the dielectric-loaded metal waveguide design for simultaneous acceleration and bunch compression, achieving a 50 times (100 fs 1.6 pC electron bunch compressed to 2 fs over an interaction distance of about 18 mm) and 62 times (100 fs to 1.61 fs over an interaction distance of 42 cm) compression for 1 MeV and 10 MeV electron bunches, respectively.

## 2. Relativistic electrodynamics in a waveguide and simulation algorithms

This section introduces the equations governing the behavior of an electron bunch in the vacuum-filled core of a waveguide, and discusses our approach in modeling this behavior. The electron bunch is made up of $N$ interacting electrons that may be modeled classically as $N$ point charges propagating according to Newton's second law:

$$\frac{d\vec{p}_i(t)}{dt} = \vec{F}_i^{\,d}(t) + \sum_{\substack{j=1,\\i\neq j}}^{N}\left(\vec{F}_{i,j}^{\,pp}(t) + \vec{F}_i^{\,wf}(t) + \vec{F}_i^{\,rr}(t)\right), \text{ with } i = 1,...,N, \qquad (1)$$

where $\vec{p}_i(t) \equiv \gamma_i(t)m\vec{v}_i(t)$ is the momentum of electron $i$ at time $t$, with $m$, $\vec{v}_i$, and $\gamma_i \equiv 1/\sqrt{1-\beta_i^2}$ being its rest mass, velocity and Lorentz factor, respectively. $\beta_i \equiv |\vec{\beta}_i|$, $\vec{\beta}_i \equiv \vec{v}_i/c$ and c is the speed of light in vacuum.

According to Eq. (1), each electron $i$ is subject to four kinds of forces: the force $\vec{F}_i^{\,d}$ exerted by the driving electromagnetic field, the sum of forces $\vec{F}_{i,j}^{\,pp}$ exerted directly by other electrons $j$, the force $\vec{F}_i^{\,wf}$ exerted by wakefields that result from electromagnetic fields of other electrons reflecting off the waveguide walls, and finally the radiation reaction force $\vec{F}_i^{\,rr}$ that the electron experiences as a result of its own radiation. In this study, we neglect $\vec{F}_i^{\,wf}$ because the relatively short propagation distances and bunch lengths make the effect of wakefields negligible. For acceleration studies involving long propagation distances, or multiple bunches of substantial charge, wakefields should be taken into consideration by implementing formulas derived in previous studies [21]. We also neglect the radiation reaction force since the employed scheme accelerates the electrons primarily via the z-directed component of the electric field, with minimal transverse wiggling. Consequently, radiation losses are negligible. Electrodynamic studies in which the radiation reaction force plays a significant role have commonly employed the Landau-Lifshitz formula [22] for the force.

The force $\vec{F}_i^{\,d}$ exerted by the driving field on electron $i$ is given by the Lorentz force equation:

$$\vec{F}_i^d(t) = q\left[\vec{E}_d(t,\vec{r}_i(t)) + \vec{v}_i(t) \times \vec{B}_d(t,\vec{r}_i(t))\right], \quad (2)$$

where $q$ is the electron's charge and $\vec{r}_i$ its position. $\vec{E}_d(t,\vec{r})$ and $\vec{B}_d(t,\vec{r})$ are respectively the electric field and magnetic flux density of the driving field. Similarly, we write the force $\vec{F}_{i,j}^{pp}$ that electron $j$ exerts on electron $i$ as

$$\vec{F}_{pp;i,j}(t) = q\left[\vec{E}_j(t,\vec{r}_i(t)) + \vec{v}_i(t) \times \vec{B}_j(t,\vec{r}_i(t))\right], \quad (3)$$

where $\vec{E}_j(t,\vec{r})$ and $\vec{B}_j(t,\vec{r})$ are respectively the electric field and magnetic flux density due to electron $j$. These fields are derived by solving Maxwell's equations for a moving point charge in vacuum via the Liénard-Wiechert potentials and the resulting electromagnetic fields are [23]

$$\vec{E}_i(t,\vec{r}) = \frac{q}{4\pi\varepsilon_0}\frac{1}{\eta_{i,\tilde{t}_i}^3(\vec{r})R_{i,\tilde{t}_i}(\vec{r})}\left\{\frac{\vec{u}_{i,\tilde{t}_i}(\vec{r})}{\gamma_i^2(\tilde{t}_i)R_{i,\tilde{t}_i}(\vec{r})} + \frac{1}{c}\left[\hat{i}_{i,\tilde{t}_i}(\vec{r}) \times \left(\vec{u}_{i,\tilde{t}_i}(\vec{r}) \times \frac{\dot{\vec{v}}_i(\tilde{t}_i)}{c}\right)\right]\right\}$$

$$\vec{B}_i(t,\vec{r}) = \frac{1}{c}\left(\hat{i}_{i,\tilde{t}_i}(\vec{r}) \times \vec{E}_i(t,\vec{r})\right), \quad (4)$$

where $\varepsilon_0$ is the permittivity of free space, $\dot{\vec{v}}_i$ the acceleration of particle $i$, $R_{i,t}(\vec{r}) \equiv |\vec{r}-\vec{r}_i(t)|$, $\hat{i}_{i,t}(\vec{r}) \equiv (\vec{r}-\vec{r}_i(t))/R_{i,t}$, $\vec{u}_{i,\tilde{t}_i}(\vec{r}) \equiv \hat{i}_{i,\tilde{t}_i}(\vec{r}) - \vec{v}_i(\tilde{t}_i)/c$ and $\eta_{i,\tilde{t}_i}(\vec{r}) \equiv dt/d\tilde{t}_i = 1 - \hat{i}_{i,\tilde{t}_i}(\vec{r}) \cdot \vec{v}_i(\tilde{t}_i)/c$. $\tilde{t}_i = \tilde{t}_i(t,\vec{r})$ is the retarded time along particle $i$'s trajectory corresponding to time $t$ and observation point $\vec{r}$. Given $t$ and $\vec{r}$, the retarded time $\tilde{t}_i$ solves the implicit equation

$$\tilde{t}_i = t - \frac{R_{i,\tilde{t}_i}(\vec{r})}{c}. \quad (5)$$

If $\vec{F}_i^d$ is the only non-zero term on the right-hand side of Eq. (1), the equation is simply an ordinary differential equation. With inter-particle interaction described by Eq. (3) and Eq. (4), the right-hand side of Eq. (1) becomes a function of $\tilde{t}_i$ as well as $t$, and the equation is no longer an ordinary differential equation. Note that Eq. (4) considers both the velocity field (near-field) and the radiation field (far-field), which are given by the first and second term respectively. If the effect of the radiation field is insignificant and we assume that each particle always travels at its current velocity during each time step, Eq. (4) can be simplified to a function of only $t$, making Eq. (1) an ordinary differential equation and reducing the computation of inter-particle forces considerably. The formulas that should replace Eq. (4) are then the space-charge formulas obtained by Lorentz-boosting the Coulomb field of each electron from the electron's rest frame to the lab frame. These formulas are used in particle tracer programs like the General Particle Tracer (GPT) [24].

We chose not to use externally-provided software packages in part to ascertain, by implementing Eq. (4), the significance of non-uniform motion and electron radiation in inter-particle interaction. It turns out that for the regime investigated in this paper, the use of the exact formulas in Eq. (4) affects overall acceleration and bunch compression results negligibly, and for computational efficiency one may simply revert to the Lorentz-boosted Coulomb fields in modeling inter-particle interaction.

We solve Eq. (1) using a fifth-order Runge-Kutta algorithm with adaptive step-size [25]. If the exact inter-particle fields in Eq. (4) are used, we adapt the Runge-Kutta algorithm to the problem by maintaining a history of $\vec{r}_i$ and $\vec{p}_i$, $i=1,\ldots,N$, in a ring buffer. At each time $t$, cubic spline interpolation is applied to compute the retarded time Eq. (5) needed in Eq. (4). Gaussian-distributions of electrons in 6-dimensional phase space are generated by applying the Box-Muller transformation to the normalized output of the *rand*() function in C, and computations of variance and covariance (required for emittance calculations) are performed using the corrected two-pass algorithm [26]. Multi-core processing capabilities are implemented using OpenMP.

In this study, we are interested in simulating bunches on the order of pCs and tens of pCs, implying that we deal with $10^7 - 10^8$ electrons. To speed up the computational process, each particle $i = 1,\ldots, N$ is treated as a macro-particle – with the charge and mass of a large number of electrons – instead of a single electron. We can verify that this approach is a good approximation if the solution converges as the number of macro-particles increases while the total number of electrons is kept constant. We have verified this for all results presented in this paper.

## 3. The pulsed TM$_{01}$ mode in a dielectric-loaded metallic waveguide

We are interested in obtaining an analytical expression that models a coherent THz pulse in the waveguide. This involves integration over the continuous-wave (CW) solutions of the waveguide. The method we use to obtain these solutions is very similar to that detailed in [27] for the optical Bragg fiber, so we only give an overview of the method here. For a general multilayer cylindrical waveguide, the continuous-wave solutions are obtained by solving the Helmholtz equation in cylindrical coordinates [23]:

$$\left(\nabla^2 + k^2\right)\begin{Bmatrix}E_z^{CW}\\H_z^{CW}\end{Bmatrix} = 0 \quad \Rightarrow \quad \frac{1}{r}\frac{\partial}{\partial r}\left(r\frac{\partial}{\partial r}\begin{Bmatrix}\psi_e\\\psi_h\end{Bmatrix}\right) + \left(k^2 - \kappa^2 - \frac{l^2}{r^2}\right)\begin{Bmatrix}\psi_e\\\psi_h\end{Bmatrix} = 0, \qquad (6)$$

where $k \equiv \omega/c = 2\pi/\lambda$, $\omega$ being angular frequency, c the speed of light in vacuum and $\lambda$ the vacuum wavelength. $E_z^{CW} \equiv \psi_e(r)\exp(i(\omega t - \kappa z \pm l\phi))$ and $H_z^{CW} \equiv \psi_h(r)\exp(i(\omega t - \kappa z \pm l\phi))$ are the complex CW z-directed electric and magnetic fields respectively, $\kappa$ is the propagation constant, $r$ the radial coordinate, $\phi$ the azimuthal coordinate, $z$ the direction of propagation along the waveguide, and $l$ a non-negative integer that determines the order of azimuthal variation. According to Eq. (6), a general solution for $\psi_e$ in layer $i$ of an $n$-layer cylindrical waveguide (the core counts as layer 1) is

$$\psi_{e;i}(r) = A_{e;i}J_l(h_ir) + B_{e;i}Y_l(h_ir), \quad r_{i-1} \leq r < r_i, \quad i = 1,\ldots,n, \qquad (7)$$

where $r_0 \equiv 0$, $r_n \equiv \infty$, and $r_i$ for $0 < i < n$ is the radial position of the boundary between layers $i$ and $i+1$. $J_l$ and $Y_l$ are Bessel functions of the first and second kind respectively, $A_{e;i}$ and $B_{e;i}$ are constant complex coefficients within each layer and $h_i \equiv (\varepsilon_{r;i}(\lambda)\mu_{r;i}(\lambda)k^2 - \kappa^2)^{1/2}$, $\varepsilon_{r;i}$ and $\mu_{r;i}$ being the dispersive relative permittivity and permeability respectively of the dielectric in layer $i$. The general solution for $\psi_{h;i}$ is identical in form to Eq. (7) except that "e" should be replaced by "h" in all subscripts. In the core, it is usually expedient to express Eq. (7) using the modified Bessel function of the first kind $I_l$, whereas in the final layer (which extends to infinity), it is usually expedient to express Eq. (7) using either the modified Bessel function of the second kind $K_l$ for confined modes or the Hankel function of the second kind $H_l^{(2)}$ for leaky modes. These functions are all exactly represented by Eq. (7) if we allow the coefficients and arguments of $J_l$ and $K_l$ to take on complex values.

The transverse electromagnetic fields are obtained from the expressions for $E_z$ and $H_z$ via Ampere's law and Faraday's law. By matching boundary conditions among adjacent dielectric layers (continuity of $E_z$, $H_z$, $E_\phi$, $H_\phi$), we obtain a characteristic matrix which has a non-trivial nullspace (zero determinant) if and only if a solution to Eq. (6) exists. Given $l$, along with the dimensions and dielectric properties of the waveguide layers, we determine numerically the set of values $\{k, \kappa\}$ for which the characteristic matrix has a zero determinant. This set of values $\{k, \kappa\}$ constitute the dispersion curves of the waveguide for a mode of azimuthal order $l$, and the $4n$ coefficients $A_{e;i}$, $B_{e;i}$, $A_{h;i}$ and $B_{h;i}$, $i=1,\ldots,n$, are the components of a $4n$-long vector in the corresponding nullspace. Up to this point, our procedure is very similar to that detailed in [27], and we direct the reader there for more information. The real-valued $z$-directed electric field $E_{z;i}$ of a pulse in any layer $i$ is constructed by an inverse Fourier transform:

$$\begin{aligned}
E_{z;i}(l,t,r,z,\varphi) &= \mathrm{Re}\left\{\int_{-\infty}^{\infty} F(\omega) E_{z;i}^{\mathrm{CW}}(l,\omega,t,r,z,\varphi)\,d\omega\right\}, \\
E_{r;i}(l,t,r,z,\varphi) &= \mathrm{Re}\left\{\int_{-\infty}^{\infty} F(\omega) E_{r;i}^{\mathrm{CW}}(l,\omega,t,r,z,\varphi)\,d\omega\right\}, \quad r_{i-1} \leq r < r_i,\ i=1,\ldots,n, \quad (8) \\
H_{\varphi;i}(l,t,r,z,\varphi) &= \mathrm{Re}\left\{\int_{-\infty}^{\infty} F(\omega) H_{\varphi;i}^{\mathrm{CW}}(l,\omega,t,r,z,\varphi)\,d\omega\right\},
\end{aligned}$$

where $F(\omega)$ is the complex envelope in the frequency domain. In Eq. (8), the same inverse Fourier transform is also applied to field components $E_r$ and $H_\phi$ to obtain their real-valued pulsed versions.

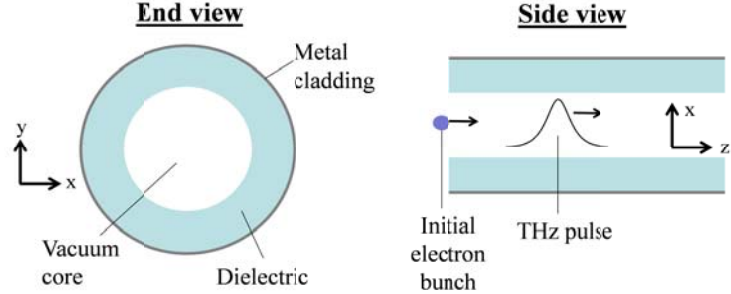

Fig. 2. Schematic of proposed waveguide and simulation setup. In this study, the dielectric is diamond ($\varepsilon_r = 5.5$). The initial relativistic electron bunch is shot through the THz pulse, which propagates at a non-relativistic group velocity.

The structure we consider in this paper is a vacuum core with a single layer of dielectric of relative permittivity $\varepsilon_r = 5.5$ (a candidate for such a dielectric is diamond [20]) with an external copper coating (Fig. 2). The spatial mode of interest is the $TM_{01}$ mode (i.e. $l = 0$ and radial variation is of the lowest order), for which only the $E_z$, $E_r$ and $H_\phi$ field components exist. The $E_z$ field peaks on axis whereas the transverse fields vanish, so an electron bunch concentrated at the waveguide axis will experience forces mainly along the direction of propagation. This facilitates longitudinal compression and acceleration of the bunch without significant transversal wiggling, which is undesirable since it tends to increase radiative losses.

To excite the $TM_{01}$ mode of the cylindrical waveguide, it would be necessary to apply a radially-polarized (preferably $TM_{01}$) beam to the waveguide. Studies on coupling linearly-polarized THz pulses into cylindrical metal waveguides show that the dominant modes excited are the $TE_{11}$, $TE_{12}$ and $TM_{11}$ modes [28], so a linearly-polarized incoming beam is unlikely to serve our purpose. Although THz pulses generated by optical rectification are typically linearly-polarized, the direct generation of radially-polarized THz pulses has been demonstrated [29, 30]. Alternatively, a scheme to convert linearly-polarized THz pulses into radially-polarized pulses may be adopted [31].

Equation (8) provides a rigorous way to compute the electromagnetic field at any point in space and time required for an electrodynamic simulation. However, performing a summation over a large number of frequency components at every time step for every macro-particle is computationally expensive. To obtain an analytical approximation for more efficient numerical simulation, notice that in the vacuum-filled core, the CW $TM_{01}$ mode is of the form:

$$E_{z,1}^{CW} = A_{e;1} I_0(q_1 r) e^{i(\omega t - \kappa z)}, \quad E_{r,1}^{CW} = A_{e;1} \frac{i\kappa}{q_1} I_1(q_1 r) e^{i(\omega t - \kappa z)}, \quad H_{\varphi,1}^{CW} = \frac{k}{\eta_0 \kappa} E_{r,1}^{CW}, \quad (9)$$

where $q_i \equiv (\kappa^2 - \varepsilon_{r;i}(\lambda)\mu_{r;i}(\lambda)k^2)^{1/2}$ is the radial wavevector and $\eta_0$ is the vacuum impedance. $I_0$ and $I_1$ are the modified Bessel functions of the first kind, of order 0 and 1 respectively. We need to make three more assumptions in the remainder of the formulation: Firstly, variations in propagation constant $\kappa$ across the spectrum are small enough that their effects on magnitude can be ignored. Secondly, variations in $\kappa$ are negligible above the second order. Thirdly, the imaginary part of $\kappa(\omega)$ is negligible beyond its $0^{th}$ order term, and the quadrature term produced by this imaginary part in $E_r$ does not contribute significantly to the field. Hence, Taylor-expanding $\kappa(\omega)$ about central angular frequency $\omega_0$ we have $\kappa(\omega_0 + \Delta\omega) \approx \kappa_0 - i\alpha + \kappa_1 \Delta\omega + \kappa_2(\Delta\omega)^2/2$, where $\kappa_i$ denotes the real part of the $i^{th}$ derivative of $\kappa(\omega)$ with respect to $\omega$ at $\omega = \omega_0$. $\alpha > 0$ to be physically valid and represents field attenuation per unit distance.

To obtain the approximate analytical field solution, the rightmost expressions of Eq. (9) should be inserted into Eq. (8). Assuming a transform-limited Gaussian pulse at $z = 0$, we have $E_z(z = 0, t) \sim \exp(-(t/T_0)^2/2)\exp(i\omega_0 t)$, where $\omega_0 \equiv k_0 c$ is the central frequency and $T_0$ is the half-width at 1/e intensity, related to the full-width-at-half-maximum (FWHM) intensity $T_{FWHM}$ as $T_{FWHM} = 2(\ln 2)^{1/2} T_0 \approx 1.665 T_0$. This is related to the spectral FWHM intensity width $\Delta\omega_{FWHM}$ as $\Delta\omega_{FWHM} = 4\ln 2/T_{FWHM}$. Finally, we have

$$E_{z;1}(t,z) \approx \text{Re}\left\{ I_0(q_{1,0}r) e^{i(\omega_0 t - \kappa_0 z)} \int_{-\infty}^{\infty} A_0 e^{-\frac{(\Delta\omega T_0)^2}{2}} e^{-i\left(\kappa_1 \Delta\omega + \frac{\kappa_2}{2}(\Delta\omega)^2\right)z} e^{i\Delta\omega t} d\Delta\omega \right\}$$

$$= \frac{|E_{z0}| I_0(q_{1,0}r)}{\left[1+\left(\kappa_2 z/T_0^2\right)^2\right]^{1/4}} e^{-\frac{(t-\kappa_1(z-z_i))^2}{2T_0^2\left[1+\left(\kappa_2 z/T_0^2\right)^2\right]}} e^{-\alpha(z-z_s)} \cos(\psi_T), \quad z \geq z_s, \quad (10)$$

where $A_0$ is an arbitrary complex constant and its role is replaced in the second line of Eq. (10) by $|E_{z0}|$, which represents the amplitude of the z-directed field at $t = 0$ and $z = z_i = z_s$, with $z_i$ being the initial position of the pulse peak. The precise relationship between $|E_{z0}|$ and the total pulse energy is complicated and must be obtained by integrating over the Poynting vector in both core and cladding regions. $q_{1,0}$ is $q_1$ evaluated at $\omega_0$. $z_s$ is the position of the start of the waveguide, where pulse attenuation begins, and before which Eq. (10) does not apply. Note that setting $z_s \neq 0$ implies that some special pulse, not transform-limited, is being coupled into the waveguide. We set $z_s = 0$ for all simulations in this paper. The carrier phase $\psi_T$ is given by

$$\psi_T = \omega_0 t - \kappa_0 z + \frac{\left(t - \kappa_1(z - z_i)\right)^2 \kappa_2 z / T_0^2}{2T_0^2 \left[1 + \left(\kappa_2 z / T_0^2\right)^2\right]} - \mathrm{atan}\left(\frac{\kappa_2 z}{T_0^2}\right) + \psi_0, \quad (11)$$

where $\psi_0$ is a real phase constant. The corresponding $E_z$, $E_r$ and $H_\phi$ fields are approximated as

$$E_{r;1}(t,r,z) \approx -\frac{\kappa_0}{q_{1,0}} \frac{\mathrm{I}_1(q_{1,0}r)}{\mathrm{I}_0(q_{1,0}r)} E_{z;1}(t,z) \tan(\psi_T), \quad H_{\phi;1}(t,r,z) \approx \frac{k_0 E_{r;1}(t,r,z)}{\kappa_0 \eta_0}. \quad (12)$$

where $k_0 \equiv \omega_0/c$. Essentially, Eq. (10) and Eq. (12) furnish an approximate analytical description of a TM$_{01}$ pulse moving with an approximate phase velocity and group velocity of $v_{ph} = \omega_0/\kappa_0$ and $v_g = 1/\kappa_1$ respectively in the vacuum core of a cylindrical waveguide. If $z_s = 0$, the pulse at the start of the waveguide ($z = z_s = 0$) is a transform-limited pulse with a peak longitudinal electrical amplitude of $|E_{z0}|$. The primary reason for introducing $z_i$ in our formulas is to control when the pulse arrives at the start of the waveguide without having to compromise the intuitive convention of having t = 0 as the initial time (when the simulation begins and the initial electron bunch starts evolving according to Eq. (1)).

## 4. Acceleration of 1.6 pC electron bunches

*4.1 Optimization procedure and acceleration results*

In this section, we optimize the dielectric-loaded metal waveguide for electron bunch acceleration and perform a rudimentary thermal damage and dielectric breakdown analysis to verify the realism of the scheme. We numerically demonstrate the acceleration of a 1.6 pC electron bunch from a kinetic energy of 1 MeV to one of 10 MeV, using a 20 mJ 10-cycle pulse centered at 0.6 THz. Note that for a 10-cycle pulse, $\Delta\omega_{FWHM}/\omega_0 = 4\ln 2/(\omega_0 T_{FWHM}) = 4\ln 2/(2\pi 10) = 4.41\%$. As will be seen in the results, some longitudinal compression is also inadvertently achieved in the process.

Optimizing the dielectric-loaded metallic waveguide for bunch acceleration involves adjusting a large number of parameters, including operating frequency, choice of waveguide mode, waveguide dimensions, THz pulse energy and pulse duration, the type of dielectric, the type of external conductor and initial electron bunch properties. To make this optimization tractable, we fix all parameters in advance based on the available technology except for three degrees of freedom: *i*) the carrier-envelope phase $\psi_0$, *ii*) the initial position of the pulse $z_i$ (with initial position of electron fixed at $z = 0$), and *iii*) the radius of the vacuum core $r_1$. In particular, we fix the phase velocity at $v_{ph} = c$ and the center frequency at $f_0 = 0.6$ THz, which limits the dielectric thickness $d$ to specific values depending on $r_1$. However, because acceleration results can be very sensitive to small variations in the value of $v_{ph}$, we take the liberty of treating $v_{ph}$ as an optimization parameter (but ensuring that $v_{ph} \approx c$) *after* using $v_{ph} = c$ to determine properties of the TM$_{01}$ waveguide mode. Therefore, four degrees of freedom are ultimately considered. In practice, after the waveguide has been fabricated according to the optimal specifications, the operating frequency should be perturbed to vary the phase velocity until maximum electron acceleration is achieved. As long as the perturbation is small, the waveguide properties should be very close to those determined for $v_{ph} = c$ and $f_0 = 0.6$ THz. The electron acceleration process is much more sensitive to small variations in $v_{ph}$ than to small variations in any other parameter caused by perturbing the operating frequency alone.

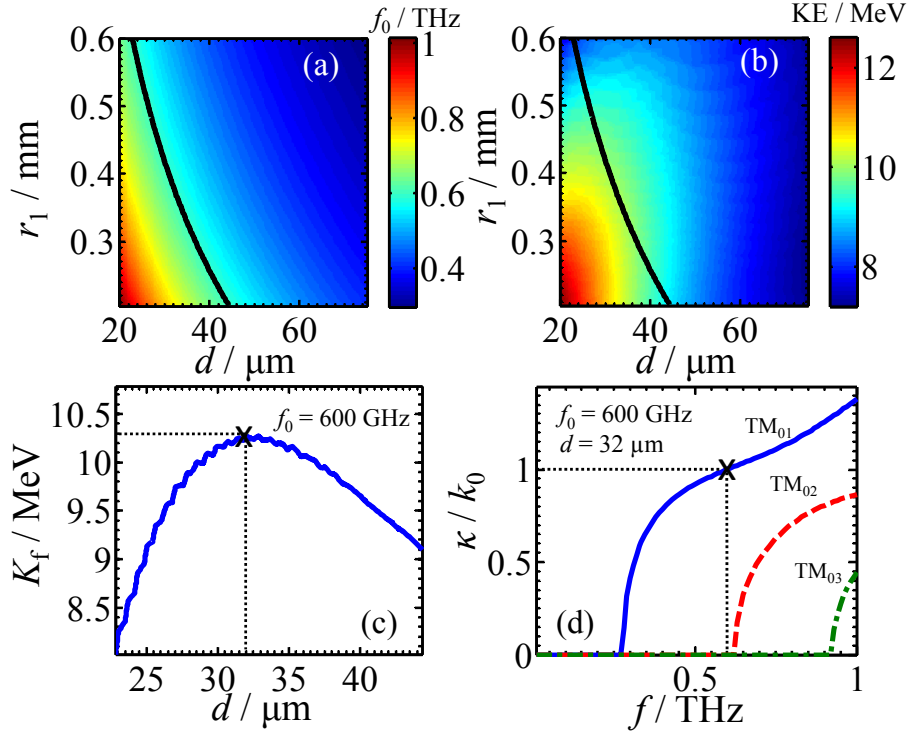

Fig. 3. Determination of the optimal waveguide for electron acceleration: (a) Color map of operation frequency $f_0$ as a function of core radius $r_1$ and dielectric thickness $d$, (b) Color map of final kinetic energy of a single electron of initial kinetic energy 1 MeV, optimized over $\psi_0$ and $z_i$, as a function of $r_1$ and $d$. The black line in (a) and (b) correspond to an operation frequency of 0.6 THz. The value of the color map in (b) along the 0.6 THz operation line is plotted in (c), where the optimum value of $d$ is identified as $d = 32$ μm. (d) The dispersion curves corresponding to the final waveguide design.

Figure 3(a) shows a color map of the operation frequency as a function of $r_1$ and $d$. As noted before, we define the operation frequency as the frequency of the $TM_{01}$ mode in the waveguide corresponding to $v_{ph} = c$. Fig. 3(b) shows a color map of the final electron kinetic energy of a single electron of initial kinetic energy 1 MeV, optimized over $\psi_0$, $z_i$ and $v_{ph}$ (ensuring that $v_{ph} \approx c$), as a function of $r_1$ and $d$. We see that greater electron acceleration is generally achieved at higher operation frequencies. However, choosing a very small wavelength makes it challenging to accelerate a large number of electrons due to smaller waveguide dimensions. As pointed out previously, the emergence of promising techniques to generate radiation in the vicinity of 0.6 THz [13] encourages us to make that choice of frequency, which has been marked out by the black contour line in Fig. 3(a). The same line is drawn in Fig. 3(b), and the optimized final kinetic energy, read along that line, is reproduced in Fig. 3(c), where an optimal choice of $d = 32$ μm, corresponding to a vacuum core radius of $r_1 = 380$ μm, is evident. In Fig. 3(d), we plot the dispersion curves corresponding to the waveguide with $d = 32$ μm, $r_1 = 380$ μm, to show that at the operating frequency, the $TM_{01}$ dispersion curve of our waveguide design is sufficiently linear within the 4.41% intensity FWHM spectral bandwidth. Hence, the electromagnetic fields are well approximated using Eq. (10) and Eq. (12).

The parameters of the final waveguide design are $d = 32$ μm, $r_1 = 380$ μm, $v_{ph} = 0.99c$, $v_g = 0.7c$, $\alpha = 5.21$/m, $T_{FWHM} = 16.7$ ps, $\kappa_2 = 4.54 \times 10^{-22}$ s$^2$/m. The 20 mJ pulse yields a $|E_{z0}|$ of about 0.9 GV/m. The initial parameters of the 1.6 pC, 1 MeV electron bunch with which we will demonstrate the acceleration are $\sigma_x = \sigma_y = \sigma_z = 30$ μm (a 100 fs bunch), $\sigma_{\gamma\beta x} = \sigma_{\gamma\beta y} = \sigma_{\gamma\beta z} = 0.006$, where $\sigma_{\gamma\beta x}$, for instance, denotes the standard deviation of $\gamma\beta_x$. Producing a 1.6 pC, 100 fs electron bunch would be a challenge for typical RF guns, but strides are being made to realize a photocathode RF gun capable of delivering the bunch we have assumed as our input [32]. Although a thorough examination of how performance is impacted by variations in the initial electron bunch properties is beyond the scope of this paper, we expect the results to deteriorate with a larger initial energy spread. 10000 macro-particles, Gaussian-distributed in every dimension of phase space, were employed in the simulation.

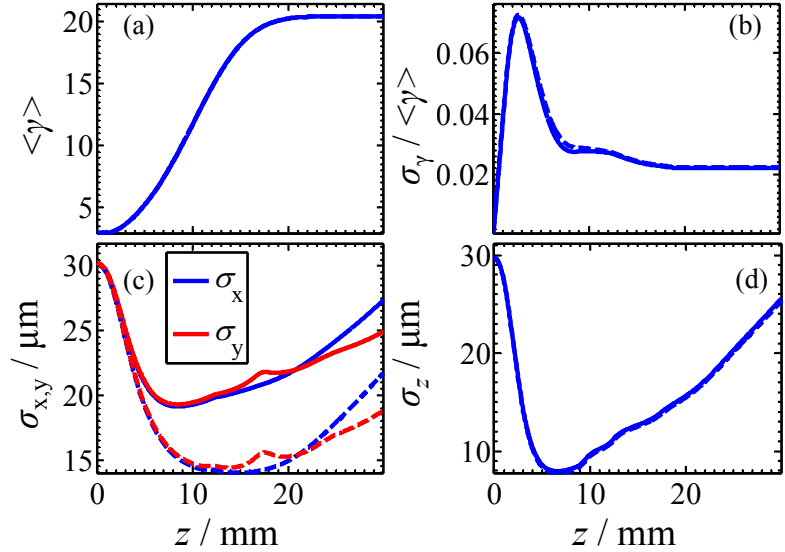

Fig. 4. Evolution of bunch parameters with mean bunch position for acceleration of a 1.6 pC electron bunch from 1 MeV to 10 MeV (kinetic energy) in about 20 mm: (a) normalized mean energy, (b) relative energy spread, (c) transverse spread and (d) longitudinal spread. The symbol $\sigma$ stands for the standard deviation of the variable in the subscript. Solid and dashed lines correspond to simulations with and without space charge respectively. 10000 macro-particles are used for the simulations. $\psi_0 = 1.34\pi$ and $k_0 z_i = 10.96\pi$. A 20 mJ, 10-cycle (16.7 ps), 0.6 THz-centered pulse is considered.

Figure 4 shows the evolution of bunch parameters as a function of mean particle position. We see from Fig. 4(a) that the 1.6 pC-bunch is accelerated from 1 MeV to 10 MeV of kinetic energy in about 20 mm, without any of its other properties deteriorating prohibitively. The corresponding average acceleration gradient is about 450 MeV/m. Note from Figs. 4(b)-(d) that, depending on the extraction point, the final bunch can possess a smaller transverse and longitudinal spread compared to the initial distribution, but the final energy spread is degraded from the initial spread.

*4.2 Injection point considerations*

In our analysis, we have assumed the freedom to inject the electron bunch into any point of the electromagnetic field. According to our computations, the optimum injection point for the electron bunch is a point within the pulse (albeit in its tail). This may be challenging to realize if both the electron bunch and the electromagnetic pulse enter the waveguide from vacuum. The objective of this section is to consider injection of the electron bunch at a point with negligible electric field values and assess the amount by which our predictions would change. The optimum THz waveguide for this case is a waveguide with $r_1 = 338$ μm and $d = 33$ μm. In addition, $v_{ph} = 0.981c$, $\psi_0 = 1.49\pi$, $k_0 z_i = 137.73$. We ensure that the electric field's amplitude at the injection point is negligible by making the amplitude $7.4 \times 10^{-15} |E_{z0}|$. The evolution of the electron bunch is shown in Fig. 5, where we observe a final kinetic energy of 8.4 MeV (instead of the 9 MeV observed before). The smaller energy gain in this case is partly due to the dispersion and attenuation that the pulse suffers from before the injected bunch begins interacting with the pulse. A final energy close to what is predicted in the previous section should therefore be achievable if the electron bunch and THz pulse can interact before the pulse has travelled too far along the waveguide.

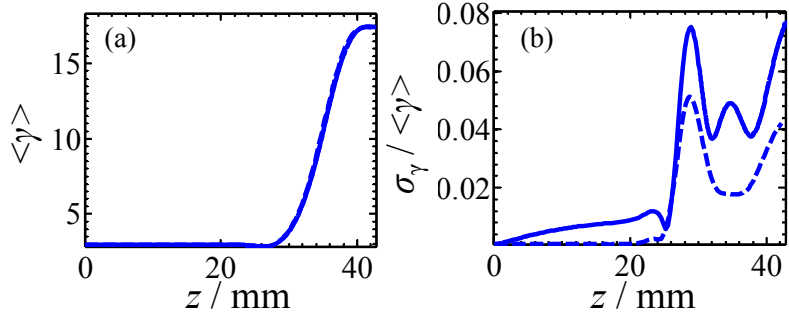

Fig. 5. Evolution of bunch parameters with mean bunch position for acceleration of a 1.6 pC electron injected at a distant point from the THz pulse peak: (a) normalized mean energy, (b) relative energy spread. Solid and dashed lines correspond to simulations with and without space charge respectively. 10000 macro-particles were used for the simulations.

*4.3 Thermal damage and dielectric breakdown considerations*

In this section, we assess the feasibility of the scheme from Sec. 4.1 in terms of its thermal damage and dielectric breakdown prospects. One concern is that the high energy injected into the waveguide and consequent energy dissipation would raise temperature of the copper coating beyond its melting point. Another concern is dielectric breakdown due to the high electric field values in the dielectric.

The energy $dG$ transferred to a differential segment of copper at position $z$ ($z = 0$ being the start of the waveguide) is related to the associated temperature rise $\Delta\theta = \theta - \theta_0$ as

$$dG = dm_{Cu} C \Delta\theta. \qquad (13)$$

The differential mass $dm_{Cu} = \rho_{Cu}(2\pi r_2 \delta_s)dz$, where $\rho_{Cu}$ is the density of copper and $\delta_s$ is the skin depth. $\theta_0$ is the original temperature of the copper and $C$ its specific heat capacity. Ignoring dispersion for simplicity (and because it is negligibly small here), we write the power propagating down the waveguide, averaged over the rapid carrier fluctuations, as

$$P(t,z) \approx P_0 e^{-\frac{(t-\kappa_1(z-z_i))^2}{T_0^2}} e^{-2\alpha z}, \quad z \geq 0, \tag{14}$$

where $P_0$ is the average power that flows into the start of the waveguide when the pulse peak arrives there. Noting that $P = -dG/dt$ and that partial derivatives are relevant here because $z$ and $t$ are independent coordinates, Eq. (13) can be written as

$$-\frac{\partial P(t,z)}{\partial z} \approx \rho_{Cu} 2\pi r_2 \delta_s C \frac{\partial \theta(t,z)}{\partial t}. \tag{15}$$

Solving Eq. (13) for $\theta$ gives

$$\theta(t,z) \approx \theta_0 - \frac{P_0 e^{-2\alpha z}}{\rho_{Cu} \pi r_2 \delta_s C} \int_{-\infty}^{t} \left(\frac{\kappa_1 t}{T_0^2} - \alpha\right) e^{-\left(\frac{t}{T_0}\right)^2} dt. \tag{16}$$

$\theta(\infty,z) - \theta_0$ gives the net temperature rise after the pulse has passed entirely through point $z$:

$$\theta(\infty,z) - \theta_0 \approx \frac{P_0 \alpha T_0}{\rho_{Cu} \sqrt{\pi} r_2 \delta_s C} e^{-2\alpha z}. \tag{17}$$

$\theta(\infty,z)$ is plotted in Fig. 6(a) for $\theta_0 = 27\ ^0C$. The relevant parameters for copper at 0.6 THz are $\rho_{Cu} = 8940$ kg/m$^3$, $C = 385$ J/kg/$^0$C and $\delta_s = 0.084$ μm.

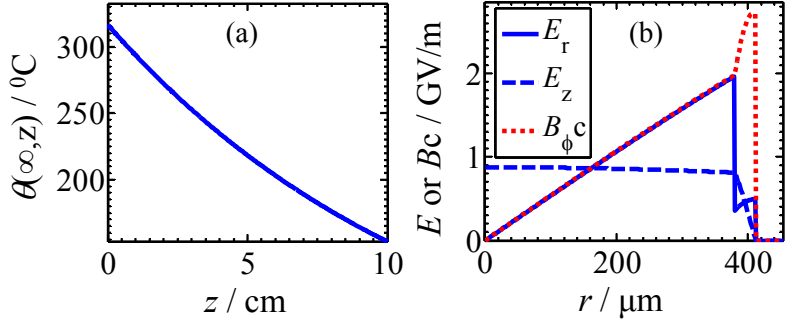

Fig. 6. Plots used to assess thermal damage and dielectric breakdown prospects for the scheme in Sec. 4.1. (a) Final temperature of copper cladding assuming initial temperature of 27 $^0$C ($z = 0$ is the start of the waveguide), and (b) Field profile of the TM$_{01}$ mode in the transverse direction of the cylindrical waveguide under study. Discontinuities occur at material boundaries (once at the vacuum-diamond interface and again at the diamond-copper interface).

From the values in Fig. 6(a), the fact that the melting point of copper is 1084 $^0$C and also that we have even neglected the conductivity of copper, we can conclude that the metal coating in the designed waveguide withstands the passage of the pulse without melting.

Figure 6(b) shows a typical profile of the electromagnetic amplitude of a mode in the transverse direction of the waveguide. The breakdown electric field for diamond has been reported as 10-20 MV/cm, depending on impurities. Reading off the plot we note that the maximum value of the electric field in the dielectric region is about 8 MV/cm. This is close to the breakdown limit though still under it, showing that it would not be feasible to enhance the performance of our design by increasing the peak power of the accelerating pulse. Since we are relatively far from the melting point, an increase in available pulse energy should be used to increase pulse duration instead of peak power.

## 5. Acceleration of 16pC and 160pC electron bunches

In this section, we explore the acceleration of electron bunches of greater charge. We see that it is feasible to use the dielectric-loaded metallic waveguide to accelerate electron bunches as large as 16 pC, but that this is not possible when the charge increases to 160 pC. All other bunch properties (including an initial kinetic energy of 1 MeV) remain the same from Sec. 4.1. We use a 20 mJ, 10-cycle, 0.6 THz-centered pulse, and the same optimized waveguide and injection conditions as in Sec. 4.1. Figure 7 shows the evolution of the electron bunch for 1.6 pC, 16 pC and 160 pC-bunches. The effects of space charge are included in all computations.

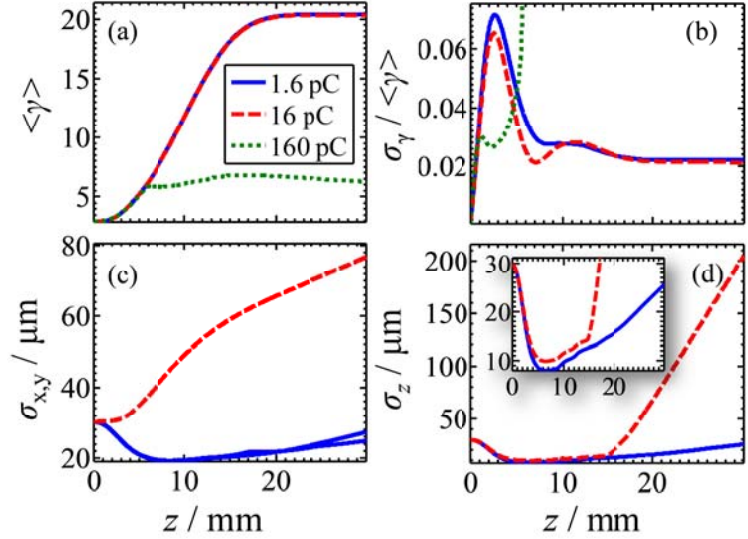

Fig. 7. Evolution of bunch parameters with mean bunch position for optimized acceleration of 1.6pC, 16pC and 160pC electron bunches: (a) normalized mean energy, (b) relative energy spread, (c) transverse spread and (d) longitudinal spread. Since the acceleration of a 160 pC bunch (green curve) is not feasible, this case is shown in (a) and (b) to demonstrate the rapidness of the Coulomb explosion, but omitted from (c) and (d) to reduce clutter. In each of the cases (1.6 pC and 16 pC) in (c), $\sigma_x$ and $\sigma_y$ practically overlap and are moreover interchangeable by a rotation of the transverse coordinates, so we do not distinguish between them, but plot both of them to show the near (but not perfect) radial symmetry in the transverse distribution. All results include space charge. 10000 macro-particles are used for all simulations. $\psi_0 = 1.34\pi$ and $k_0 z_i = 10.96\pi$. A 20 mJ, 10-cycle (16.7 ps), 0.6 THz-centered pulse is used in all cases.

From Figs. 7(a) and 7(b), we observe that there is little difference in the mean kinetic energy and energy spread evolution of a 16 pC-bunch and a 1.6 pC-bunch. The energy spread of a 160 pC-bunch, however, deteriorates prohibitively and the bunch is not significantly accelerated. Since this rules out the feasibility of accelerating a 160 pC-bunch, we have omitted its plots from Figs. 7(c) and 7(d). The inability of the waveguide to accelerate a 160 pC-bunch is due to the overriding strength of the Coulomb repulsion, driving the electrons into the walls of the waveguide before significant acceleration takes place. Figure 7(c) explains how the 1.6 pC and 16 pC-bunches are able to have such similar energy and energy spread profiles during the acceleration: the greater Coulomb repulsion in the 16 pC Coulomb is counter-balanced by larger transverse inter-particle spacing. Figure 7(d) shows that due to the larger amount of space charge, the 16 pC expands rather rapidly compared to the 1.6 pC-bunch after the pulse has slipped behind the bunch, so a 16 pC-bunch accelerated via this scheme is likely to be useful for a shorter duration after being fully accelerated.

## 6. Concurrent phase-limited compression and acceleration of 1.6pC-bunches

In this section, we optimize our waveguide design for simultaneous acceleration and bunch compression. We demonstrate phase-limited (longitudinal) bunch compression of 50 and 62 times for electron bunches of initial kinetic energy 1 MeV and 10 MeV respectively. By "phase-limited" we mean that the maximum compression results do not change substantially when space charge is removed from the simulations.

As in previous sections, we use a 20 mJ, 0.6 THz-centered pulse. For each case (the 1 MeV case and the 10 MeV case), the waveguide and injection conditions are optimized exactly as described in Sec. 4.1, except that in addition to $\psi_0$, $z_i$, $r_1$, and $v_{ph}$, we also optimize over pulse duration $T_{FWHM}$ (keeping total energy constant at 20 mJ), for a total of five optimization parameters. The initial conditions of the electron bunch, unless otherwise specified, are the same as those in Sec. 4.1.

To optimize for simultaneous acceleration and compression, the figure-of-merit found to be most useful is the ratio of energy to bunch-length of the electron bunch. Unlike in Sec. 4.1, where we optimized using a single particle, here we optimized using 100 macro-particles and included the effects of space charge. The optimized results are then verified with simulations that use 10000 macro-particles.

For the 1 MeV case, our optimized parameters are $\psi_0 = 0.73\pi$, $k_0 z_i = 13.3\pi$, $r_1 = 447$ μm, $T_{FWHM} = 13.1$ ps (7.86 cycles). The evolution of the electron bunch parameters under these optimal conditions are presented in Figs. 8(a)-8(c), where we observe a small net acceleration and a phase-limited compression of the electron bunch from 100 fs (30 μm) to about 2 fs over an interaction distance of about 18 mm. Note that there is a limited time window during which the electron bunch remains maximally compressed. Conceptually, this is unavoidable due to the presence of space charge which causes the bunch to expand after the bunch has slipped from the THz pulse.

For the 10 MeV case, our optimized parameters are $\psi_0 = 1.02\pi$, $k_0 z_i = 206\pi$, $r_1 = 597$μm, $T_{FWHM} = 170.5$ ps (102.3-cycle). The evolution of the electron bunch parameters under these optimal conditions are presented in Figs. 8(d)-8(f), where we observe a phase-limited compression of the electron bunch from 100 fs to 1.61 fs over an interaction distance of 42 cm. Although the bunch is compressed by a slightly larger factor than in the 1 MeV case, the much larger interaction distance suggests that the superior strategy to obtain a high energy, compressed bunch is to compress it before acceleration.

## 7. Conclusion

The quest to realize an efficient, practical compact accelerator for electron bunches of substantial charge will likely involve a tradeoff between the large wavelengths but low acceleration gradient of RF accelerators, and the high acceleration gradient but small wavelengths available at optical frequencies. The trade-off between acceleration gradient and wavelength, together with the emergence of efficient methods to generate coherent pulses at THz frequencies, make electron acceleration at THz frequencies a promising candidate for the substantial acceleration and compression of pico-Coulomb electron bunches. In this paper, we numerically demonstrated the acceleration of a 1.6 pC electron bunch from a kinetic energy of 1 MeV to one of 10 MeV over an interaction distance of about 20 mm, using a 20 mJ pulse centered at 0.6 THz in a dielectric-loaded metallic waveguide. We have also analyzed the implications of using an arbitrarily distant injection point, as well as the prospects of dielectric breakdown and thermal damage for our optimized design.

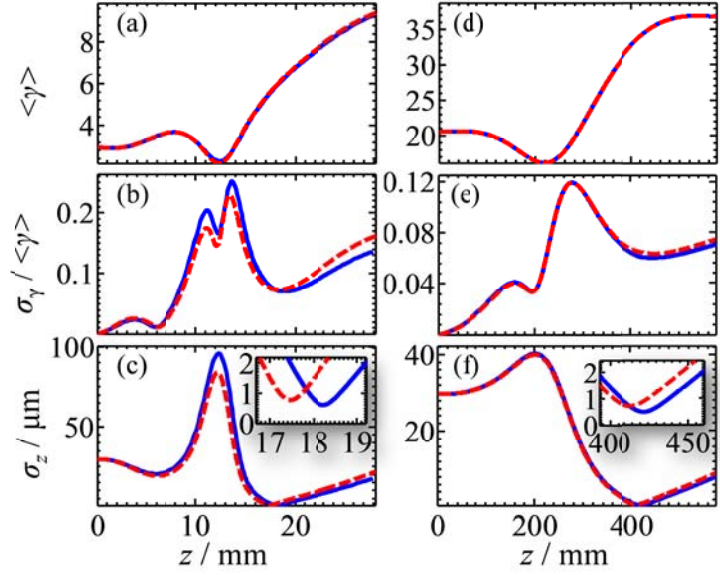

Fig. 8. Concurrent compression and acceleration of a 1.6 pC electron bunch under optimized conditions, with a compression factor of 50 and 62 achieved for initial kinetic energies of 1 MeV and 10 MeV respectively. The evolution of (a) normalized mean energy, (b) relative energy spread and (c) longitudinal spread are shown for a 1 MeV bunch subjected to a 20 mJ, 7.86-cycle (13.1 ps), 0.6 THz-centered pulse ($\psi_0 = 0.73\pi$ and $k_0 z_i = 13.3\pi$). Similarly, the evolution of (e) normalized mean energy, (f) relative energy spread and (g) longitudinal spread are shown for a 10MeV bunch subjected to a 20 mJ, 102.3-cycle (170.5ps), 0.6 THz-centered pulse ($\psi_0 = 1.02\pi$ and $k_0 z_i = 206\pi$). Blue solid curves and red dashed curves indicate simulations with and without space charge respectively. 10000 macro-particles were used for all simulations

In addition, we investigated the acceleration of 16 pC and 160 pC 1 MeV electron bunches, observing that performance does not change significantly for a 16 pC-bunch, but deteriorates prohibitively for a 160 pC-bunch due to the overwhelming Coulomb repulsion. Finally, we optimized the dielectric-loaded metal waveguide design for simultaneous acceleration and bunch compression, achieving a 50 times (100 fs 1.6 pC electron bunch compressed to 2 fs over an interaction distance of about 18 mm) and 62 times (100 fs to 1.61 fs over an interaction distance of 42 cm) compression for a 1 MeV and 10 MeV electron bunch respectively. These results were achieved with a 20 mJ coherent THz pulse centered at 0.6 THz, and encourage the exploration of coherent THz pulse-driven electron acceleration as a path to compact electron acceleration and bunch compression schemes.

## Acknowledgment

This work is supported by DARPA AXIS Program under grant N66001-11-1-4192. We thank W. S. Graves, P. Piot and A. Sell for helpful discussions. L. J. Wong acknowledges funding from the Agency for Science, Technology and Research, Singapore.